# Light Bridges and Solar Active Region Evolution Processes

Fuyu Li,[1,2,3,4] Changhui Rao,[1,4,5] Xinhua Zhao,[2] Yang Guo,[3] Xiaoying Gong,[1,5,4] Yuhao Chen,[6,5] Nanbin Xiang,[6,2] and Huaning Wang[4,7]

[1]*Institute of Optics and Electronics, Chinese Academy of Sciences, P.O. Box 350, Chengdu 610209, People's Republic of China*
[2]*State Key Laboratory of Space Weather, National Space Science Center, Chinese Academy of Sciences, People's Republic of China*
[3]*Key Laboratory of Modern Astronomy and Astrophysics of the Ministry of Education, Nanjing University, 210093, People's Republic of China*
[4]*National Laboratory on Adaptive Optics, Chengdu 610209, People's Republic of China*
[5]*University of Chinese Academy of Science, Beijing, People s Republic of China*
[6]*Yunnan Observatories, Chinese Academy of Sciences, Kunming 650011, People's Republic of China*
[7]*Faculty of Electrical Engineering and Computer Science, Ningbo University, Ningbo 315000, People's Republic of China*

## ABSTRACT

The formation mechanism of light bridges (LBs) is strongly related to the dynamic evolution of solar active regions (ARs). To study the relationship between LB formation and AR evolution phases, we employ 109 LB samples from 69 ARs in 2014 using observational data from the Helioseismic and Magnetic Imager on board the Solar Dynamics Observatory (HMI/SDO). LBs are well matched with the weak field lanes (WFLs), except that aligned on the polarity inversion line of $\delta$ sunspots. For penumbral intrusion (type-A) and umbral-dot emergence (type-C) LBs, the WFLs represent the splitting of magnetic flux systems. The sunspots tend to decay and split into several parts after type-A and type-C LBs formed. For sunspot/umbra merging (type-B) LBs, the WFLs declining are caused by collisions of flux systems. The sunspots merge and keep stable after type-B LBs formed. We conclude that type-B LBs are formed by collisions of flux systems, while type-A and type-C LBs are generated by splits. The time differences ($\delta T$) between LBs appearing and ARs peaking have average value of 1.06, -1.60, 1.82 for type-A, B, C LBs, with the standard deviation of 3.27, 2.17, 1.89, respectively. A positive value of $\delta T$ means that the LB appear after AR peaking, whereas a minus $\delta T$ before the peak. Type-A LBs trend to form in the decaying phase or around the peak time. Type-B LBs are more likely to be formed in the developing phase. Type-C LBs mostly take shape in the decaying phase of ARs.

## 1. INTRODUCTION

Light bridges (LBs) appear elongated structures within sunspots connected to penumbrae, obviously brighter than umbra in photospheric observations (Shimizu 2011; Liu 2012; Song et al. 2017; Kamlah et al. 2023). It is generally believed that the magnetic fields of LBs are weaker and more inclined than the strong-field umbra surroundings, implying un-completely suppressed convection (Leka 1997; Spruit & Scharmer 2006; Bharti 2015; Toriumi et al. 2015; Wang et al. 2018b; Yang et al. 2019; Wang et al. 2022). Shimizu et al. (2009) suggested that a twisted magnetic flux rope are trapped below the cusp-shaped magnetic field along the light bridge (Jurčák et al. 2006; Wang et al. 2018a). High resolution observations show some fine structures within the LBs, e.g. dark lanes with upflows, dark knots with downflows (Rouppe van der Voort et al. 2010; Yang et al. 2015; Zhang et al. 2018). Guglielmino et al. (2017) observed an unusual filament-like LB and regard it as a flux rope within umbra (Guglielmino et al. 2019). According to various

Corresponding author: Changhui Rao
chrao@ioe.ac.cn

Corresponding author: Fuyu Li
lifuyu@ioe.ac.cn

morphology features, LBs at least contain two categories: granular LBs and filament LBs (Rimmele 2008; Kleint & Sainz Dalda 2013; Tiwari et al. 2013; Lagg et al. 2014; Wang et al. 2018b; Guglielmino et al. 2017, 2019). Observations expose a highly dynamical scenario of sunspot LBs. There are small-scale actives above LBs, such as H$\alpha$ surges, light wall oscillations, and small-scale bright blobs ejected from the sunspot light bridge, that could sustainedly heat the chromosphere and transition region (Louis et al. 2023; Tian et al. 2018; Bai et al. 2019; Zheng et al. 2019; Li et al. 2021a; Hou et al. 2022).

Li et al. (2021b) classified LBs based on their formation processes, and present the first statistical investigation on the formation of LBs. They divide these formation processes into three categories: penumbral intrusion (type-A), sunspot merging (type-B) and umbral-dot emergence (type-C); (Li et al. 2021b). Type-A LBs are formed when some parts of penumbrae intrude towards the umbrae (Rimmele 2004; Katsukawa et al. 2007; Ortiz et al. 2010; Hou et al. 2020), which is the most common way (Li et al. 2021b). A positive correlation between the duration of the formation process and the length of LB, suggests a similar speed of penumbral intrusion in different sunspots (Li et al. 2021b). Type-B LBs are arised from the coalescence of two or more sunspots/pores (Zirin & Wang 1990; Li et al. 2021b). Furthermore, Li et al. (2021b) emphasized that umbral dots (UDs) in the center of a sunspot have potential to facilitate the formation of light bridges, categorized into type-C. Statistical results mention that most LBs are formed in ARs with complex magnetic field configuration (Li et al. 2021b).

Weak field lanes (WFLs) appear in umbrae and penumbrae of sunspots, caused by magnetic flux motions driven by flows in the convection zone, and matched with both strong LBs and faint LBs (Wang et al. 2022). Quasi separatrix layers (QSLs) can be computed by force-free field extrapolation (Yan & Sakurai 2000; Wang 1997; Wang et al. 2001), that are different from WFLs in the dominant sunspot but have possibility to partially overlap in the umbra (Wang et al. 2022). Wang et al. (2022) suggested that WFLs might be regarded as rifts among splitting flux systems, that cannot be attributed to topological evolution of magnetic fields. In addition, they point the possibility that QSLs arise from collisions of flux systems, while WFLs arise from splits of flux systems. LBs might appear in all stages of the sunspot lifetime because flux system collisions and splits could exist during the full lifetime of solar active regions (ARs). In the developing phase of sunspots, the collision process is dominant, while splits predominate in the decaying phase (Wang et al. 2022). However, there is only one case as decaying phase studied in Wang et al. (2022). Here, we tend to statistical investigate the formation mechanism of different types of LBs along with ARs evolution processes.

LBs are more likely to be formed in ARs with complicated and multipolar magnetic fields (Li et al. 2021b). Besides their local fine structures, LBs are characterized by magnetic flux systems of ARs, that would be beneficial to understand the LB formation mechanisms. Section 2 describes the observation data sources, and the characteristic parameters of LBs. Section 3.1 explains active region evolution phases. In Section 3.2 - 3.4, we present three different formation processes of LBs and analyze the relevant flux system motivations. Finally, in Section 4, the relationship between LBs and flux systems are discussed.

## 2. OBSERVATIONAL DATA

Using the JHelioviewer software (Müller et al. 2017) (also see the website at: https://www.helioviewer.org) and HMI images to select cases of LB formation, Li et al. (2021b) identified 144 isolated LBs in whole 2014, the most active year of the 24th solar cycle, and obtained their details, such as type, length, and formation process duration. The following criteria were employed for LB selection: (1) the LB should be well isolated from others; (2) the whole LB formation process should be clearly and compeletly observed (Li et al. 2021b). Table 1 lists 7 LBs appearing in AR 11944 in time sequence, with the type of each LBs in the 3rd column. Listed in the 2ed column, the formation time refers to the time when both ends of an LB get connected to the penumbra (Li et al. 2021b). Based on the characteristics of formation processes, LBs can be formed by three different ways: penumbral intrusion (type-A), sunspot/umbra merging (type-B) and umbral-dot emergence (type-C) (Li et al. 2021b). In Table 2, the 2-6th columns list the AR number, initial recorded date, incipient heliographic coordinate, last recorded date, last heliographic coordinate. Types of LBs are listed in the 7th column of Table 2 for each AR, sorted by formation time as if multiple LBs appear successively in one AR. And, the 8th column gives the formation time of the LB, or time intervals for multiple LBs in one AR. In a solar active region (AR), it is possible to exist several LBs of different type. The last column list the date to sunspot area peak for every AR. The information of ARs is published by the website of National Oceanic and Atmospheric Administration (NOAA, https://www.swpc.noaa.gov/products/solar-region-summary). Here we just pick out those LBs (from Li et al. (2021b)) with their length equal or greater than 8 arcsec, to match better with large-scale ARs. Thus, 109 LBs from 63 ARs (Table 2) are taken into account, in which 53, 45, 11 cases for type-A, B, C, respectively.

**Table 1.** Types evolution of multiple light bridges in AR 11944.

| No. | Time | Type of LBs |
|---|---|---|
| 1 | 2014/1/3 02:40 | B |
| 2 | 2014/1/4 04:40 | B |
| 3 | 2014/1/4 10:42 | B |
| 4 | 2014/1/7 14:09 | A |
| 5 | 2014/1/7 07:09 | A |
| 6 | 2014/1/9 05:49 | A |
| 7 | 2014/1/10 23:49 | C |

The Helioseismic and Magnetic Imager (HMI, Scherrer et al. 2012) on board the Solar Dynamics Observatory (SDO, Pesnell et al. 2012), provides Space-weather HMI Active Region Patches (SHARP) data series, $hmi.sharp\_cea\_720s$. The cadence is 12 minutes. The resolution is $0.''5$ per pixel. To investigate the solar active region evolution and the formation process of LBs in detail, we employ continuum images and magnetograms. Based on the three components of filed, the total magnetic strength can be computed by the following equation,

$$|B| = \sqrt{B_p^2 + B_r^2 + B_t^2}. \qquad (1)$$

where parameters $B_p$, $B_r$, $B_t$, represent parallel, radial, and transverse components of magnetic field respectively.

### 3. LIGHT BRIDGE TYPE AND ACTIVE REGION EVOLUTION PHASES

#### 3.1. *Active Region Evolution Phases*

ARs are regarded as combinations of dynamical flux systems (Jiang 2020; Jiang et al. 2023; Zhang & Jiang 2022; Cameron & Schüssler 2023). Their evolution processes generally include developing phase and decaying phase. In the decaying phase, flux systems decline and separate from each other. It is obviously observed that the sunspot splits and declines gradually in the temporal continuum maps and magnetic maps (Figures 1 and 4). In the developing phase, flux systems constantly emerge to the solar surface (Jiang 2020; Wang et al. 2021), and converge toward each other, and then collisions happen naturally. This process is shown as sunspots colliding and merging together in corresponding continuum observation (Figure 2). Flux systems collision and split might exist during the full lifetime of ARs, that is why LBs appear in all stages of the sunspot lifetime (Wang et al. 2022). In the developing phase of ARs, collisions are dominant, but splits in the decaying phase (Wang et al. 2022). Thus, we believe that type-B LBs tend to be formed in the developing phase, while type-A and type-C LBs in the decaying phase. Consequently, for multiple types of LBs in the same active region, type-B is expected prior to other types.

LBs are classic lightful structures within sunspots, indicated by red arrows in Figures 1 and 2. And, it has been known that LBs are corresponding to weak field lanes (WFLs) (Lites et al. 1991; Katsukawa et al. 2007; Li et al. 2021b; Wang et al. 2022). Logically, WFLs could take shape before flux systems gathering or after splitting. In converging process, WFLs are expected to fade away, representting flux systems merging together. In separating process, WFLs are supposed to extend gradually, signifying flux systems splitting. Considering AR evolutions with the same LB type are similar, we select several cases to investigate the evolution in the following three subsections.

#### 3.2. *Active Region Evolution with Type-A Light Bridge*

Previously, some case studies have shown that LBs could form when the penumbrae intrude inside umbrae (Katsukawa et al. 2007; Louis et al. 2020). Penumbral intrusion(type-A) is the most common process responsible for the formation of LBs, as the statistical result suggested by Li et al. (2021b). For example, Figure 1 shows the formation and subsequent evolution of a type-A light bridge within a spot in NOAA 12235. The top line shows the continuum intensity images, while the bottom line shows magnetic field strength images. We can see a bright narrow structure intruding from the penumbra towards the umbra, and gradually reaches the other side of the penumbra. Then, the new LB(indicated by a red arrow) forms at around 12:34 UT on 2014 December 16. To explore the magnetic flux systems of active regions with LBs, the magnetic field intensity are mapped in the second row of Figure 1. The WFL, corresponding to the LB, is clear in the field strength images (in Figure 1), which indicates the split of flux system. Meanwhile, another light bridge is formed in the same way in the north part of the same sunspot. It should be noted that the subsequent evolution of this sunspot after the LBs formed. The light bridges within the spot broaden, while the penumbrae in the end of LBs disappear. Then the sunspot is split into several separated parts in the following days, that means the decaying of the sunspot. In the bottom panel of Figure 1, the total magnetic flux evolution curve is monotone decreasing, that confirms the decaying. Thus, the appearance of type-A light bridge could signify the decaying phase of the local AR.

### 3.3. *Active Region Evolution with Type-B Light Bridge*

LBs can also be formed when two or more sunspots/pores merge (Zirin & Wang 1990), categorized into type-B in Li et al. (2021b). It is also a common way to form LBs (Li et al. 2021b). Figure 2 shows an example of type-B formation process in active region NOAA 12060, and the later evolution. While two sunspots close to each other, the gap between them decreases. Then, penumbrae appear and contact with the light lane at both ends, stating that the two spots completely merge into one spot. The region between the two approaching spots evolves into an LB at 07:00 UT on 2014 May 14, marked by the red arrow in Figure 2. A comparison between the continuum images and magnetic strength maps tells that the magnetic field associated with the LB is evidently weaker than that of the surrounding umbra. Then, the WFL fading and disappearing are caused by magnetic flux systems collisions. As to the subsequent temporal evolution of this sunspot after the type-B LB formed, the new merged umbra keeps nearly stable (not splitting) in the following days, and the LB gradually fade away. As shown by the flux evolution curve in Figure 2, the LB was formed after a flux increasing process. This fact suggests that type-B LB are associated with the developing phase of AR.

**Table 2.** Information for ARs and LBs therein.

| No. | Active Regions | $t_0$ of ARs | Location | $t_{end}$ of ARs | Location | Type Situation of LBs | Dates for LBs | Peak Dates for ARs |
|---|---|---|---|---|---|---|---|---|
| 1 | 11944 | 01/01 | S07E75 | 01/14 | S12W88 | B→B→B→A→A→A→C | 01/03-01/10 | 01/08 |
| 2 | 11946 | 01/04 | N12E41 | 01/13 | N08W87 | C→A | 01/09-01/10 | 01/10 |
| 3 | 11949 | 01/08 | S15E70 | 01/20 | S15W84 | A | 01/16 | 01/12 |
| 4 | 11959 | 01/18 | S25E68 | 01/30 | S23W88 | B | 01/19 | 01/21 |
| 5 | 11960 | 01/18 | S14E77 | 01/30 | S15W82 | A→C | 01/24-01/25 | 01/22 |
| 6 | 11967 | 01/27 | S14E76 | 02/09 | S13W89 | B→A→B | 01/31-02/07 | 02/04 |
| 7 | 11968 | 01/27 | N10E75 | 02/09 | N11W85 | B | 02/02 | 02/06 |
| 8 | 11973 | 02/05 | N06E63 | 02/16 | N06W87 | B | 02/07 | 02/09 |
| 9 | 11974 | 02/05 | S11E76 | 02/18 | S12W89 | B→B→A | 02/14-02/16 | 02/14 |
| 10 | 11976 | 02/08 | S15E78 | 02/21 | S13W94 | A | 02/18 | 02/10 |
| 11 | 11977 | 02/10 | S11E77 | 02/22 | S08W85 | A | 02/12 | 02/13 |
| 12 | 11982 | 02/17 | S11E70 | 03/01 | S11W86 | A | 02/27 | 02/23 |
| 13 | 11991 | 02/26 | S25E65 | 03/08 | S26W61 | B→A | 02/28-03/01 | 03/02 |
| 14 | 11996 | 03/02 | N13E55 | 03/13 | N16W90 | B | 03/11 | 03/12 |
| 15 | 12002 | 03/08 | S19E64 | 03/19 | S18W86 | A→A→A | 03/11-03/13 | 03/13 |
| 16 | 12003 | 03/10 | N06W12 | 03/16 | N05W91 | B | 03/14 | 03/13 |
| 17 | 12004 | 03/10 | S08E64 | 03/21 | S09W85 | B | 03/19 | 03/19 |
| 18 | 12010 | 03/17 | S14E64 | 03/29 | S13W90 | A→A | 03/23-03/24 | 03/24 |
| 19 | 12011 | 03/18 | S07W22 | 03/23 | S06W92 | B | 03/19 | 03/22 |
| 20 | 12014 | 03/19 | S15E74 | 03/31 | S13W91 | B | 03/20 | 03/28 |
| 21 | 12021 | 03/28 | S15E50 | 04/07 | S13W82 | B→A→B | 03/30-03/31 | 04/02 |
| 22 | 12033 | 04/09 | N12E62 | 04/20 | N11W84 | C | 04/15 | 04/11 |

| | | | | | | | |
|---|---|---|---|---|---|---|---|
| 23 | 12035 | 04/11 | S15E74 | 04/23 | S13W85 | A | 04/14 | 04/23 |
| 24 | 12036 | 04/13 | S18E25 | 04/22 | S17W92 | B | 04/15 | 04/17 |
| 25 | 12038 | 04/15 | S12E64 | 04/25 | S09W74 | B | 04/22 | 04/23 |
| 26 | 12047 | 04/26 | S19E56 | 05/06 | S17W82 | B | 05/03 | 05/02 |
| 27 | 12049 | 04/27 | S07E71 | 05/09 | S07W89 | B→A | 05/01-05/03 | 05/01 |
| 28 | 12055 | 05/05 | N10E67 | 05/17 | N13W91 | A | 05/11 | 05/09 |
| 29 | 12056 | 05/06 | N05E70 | 05/18 | N04W91 | C→A→C | 05/11 | 05/09 |
| 30 | 12057 | 05/07 | N15E65 | 05/18 | N16W81 | A | 05/12 | 05/09 |
| 31 | 12060 | 05/11 | S16E37 | 05/20 | S13W89 | A→B→A | 05/12-05/17 | 05/14 |
| 32 | 12061 | 05/11 | S24E67 | 05/23 | S25W88 | A | 05/16 | 05/15 |
| 33 | 12080 | 06/03 | S13E57 | 06/14 | S11W91 | B→B | 06/10 | 06/13 |
| 34 | 12085 | 06/06 | S21E24 | 06/14 | S19W84 | B→A | 06/09-06/11 | 06/09 |
| 35 | 12104 | 06/28 | S10E78 | 07/11 | S12W92 | A | 07/06 | 07/02 |
| 36 | 12107 | 06/29 | S20E75 | 07/12 | S19W93 | A | 07/03 | 07/02 |
| 37 | 12108 | 07/01 | S08E68 | 07/13 | S07W89 | B | 07/06 | 07/08 |
| 38 | 12119 | 07/18 | S23E16 | 07/25 | S21W82 | B | 07/20 | 07/21 |
| 39 | 12126 | 07/26 | S10E16 | 08/03 | S09W96 | B | 07/30 | 07/30 |
| 40 | 12132 | 07/31 | S18E60 | 08/10 | S19W75 | B | 08/02 | 08/07 |
| 41 | 12135 | 08/06 | N11E64 | 08/17 | N16W87 | A | 08/07 | 08/08 |
| 42 | 12144 | 08/14 | S17W19 | 08/19 | S17W87 | B | 08/16 | 08/17 |
| 43 | 12146 | 08/16 | N09E76 | 08/28 | N08W86 | A | 08/20 | 08/26 |
| 44 | 12151 | 08/23 | S07E71 | 09/04 | S08W88 | A→A | 08/25-09/01 | 08/27 |
| 45 | 12152 | 08/28 | S18E55 | 09/08 | S15W92 | B | 09/03 | 09/04 |
| 46 | 12153 | 08/29 | S11E18 | 09/06 | S09W92 | B→B→B | 09/02-09/04 | 09/04 |
| 47 | 12157 | 09/04 | S13E68 | 09/16 | S15W88 | C→A→A→A | 09/08-09/14 | 09/06 |
| 48 | 12158 | 09/04 | N16E83 | 09/17 | N15W93 | A | 09/07 | 09/11 |
| 49 | 12172 | 09/20 | S09E75 | 10/03 | S08W91 | C→A→A→A→B | 09/27-09/28 | 09/23 |
| 50 | 12175 | 09/25 | N15W12 | 10/01 | N18W90 | B→B | 09/27 | 09/29 |
| 51 | 12186 | 10/07 | S19E73 | 10/19 | S20W84 | A | 10/14 | 10/10 |
| 52 | 12192 | 10/17 | S13E68 | 10/30 | S15W94 | A→A | 10/26-10/28 | 10/26 |
| 53 | 12193 | 10/19 | N05E10 | 10/26 | N06W87 | B→C | 10/20-10/23 | 10/22 |
| 54 | 12203 | 11/01 | N12E08 | 11/08 | N13W84 | B | 11/02 | 11/03 |
| 55 | 12209 | 11/12 | S13E73 | 11/26 | S16W91 | C | 11/21 | 11/18 |
| 56 | 12216 | 11/20 | S13E68 | 12/02 | S14W90 | A→A→A | 11/23-11/28 | 11/24 |
| 57 | 12222 | 11/26 | S20E70 | 12/08 | S19W91 | A→C→A→A | 12/02-12/05 | 12/04 |
| 58 | 12227 | 12/02 | S03E71 | 12/14 | S04W87 | A | 12/07 | 12/09 |
| 59 | 12230 | 12/06 | S14E57 | 12/17 | S14W88 | B→A | 12/11-12/12 | 12/13 |
| 60 | 12234 | 12/10 | N05E18 | 12/17 | N05W84 | B | 12/12 | 12/14 |
| 61 | 12235 | 12/10 | S07E63 | 12/21 | S08W83 | A | 12/16 | 12/11 |
| 62 | 12241 | 12/14 | S11E59 | 12/25 | S08W89 | B→A | 12/16-12/17 | 12/19 |
| 63 | 12242 | 12/14 | S20E38 | 12/23 | S16W85 | B→B | 12/17 | 12/19 |

[1] Note: Type situation of LBs contains number, type and appearing time sequence. Dates for LBs refer to the appearing date for single LB, or dates for the first LB and the last LB. Peak Dates for ARs refer to the date of maximum sunspot area.

In general, type-B LB is developed by convergence of two or more spots in the same magnetic field polarity. But, there is a special case (Figure 3), as that in $\delta$ sunspot, formed by convergence of spots in opposite polarities. As we know, such light bridges stay along the polarity inversion line and harbor strong field (Livingston et al. 2006; Wang et al. 2018a). Shown in Figure 3, this LB (marked with red arrows) in NOAA 11967 is not associated with an obvious WFL in the bottom magnetic strength maps. There are some saturated pixels corresponding to the LB, because vector magnetograms from HMI on board SDO use Stokes inversion with the maximum magnetic field strength limited to 5000 G (Scherrer et al. 2012; Wang et al. 2018a). Castellanos Durán et al. (2020) confirms the strong fields in the same light bridge, by employing Hinode/BFI filtergrams and HMI data. Thus, type-B LBs are associated with weaker fields than that of surrounding umbrea, except those in $\delta$ sunspot. In addition, the active region in Figure 3 has similar developing trend after the LB formation as the case in Figure 2.

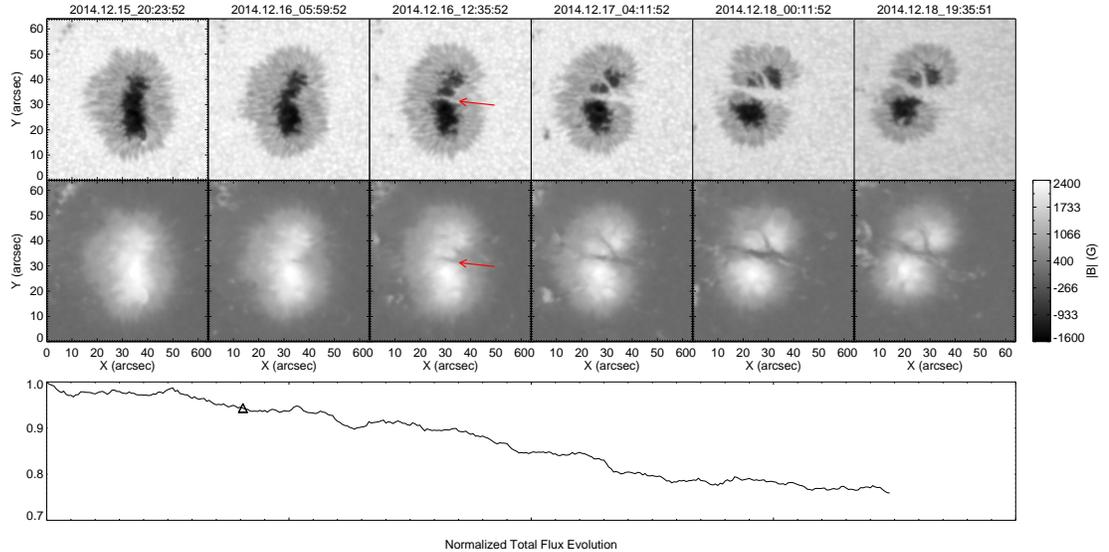

Movie 1

**Figure 1.** The formation process of a type-A (penumbral intrusion) LB and subsequent AR evolution in maps of *hmi.sharp_cea_720s* from NOAA 12235. The top row shows the continuum intensity. The second row shows the total magnetic strength. The red arrows indicate location of the newly formed LB. The normalized total magnetic flux evolution (sum total in the same filed of view as second row panels) is displayed by the curve in the bottom panel. The triangle marks the LB appering. The duration of this curve is same to the above panels. (This figure is available as an animation, in the same field of view (FOV), from 2014.12.15_18:59:52 to 2014.12.19_18:59:51, showing the evolution of the LB and sited sunspot in continuum wavelength and magnetogram.)

### 3.4. *Active Region Evolution with Type-C Light Bridge*

Light Bridge can also be evolved from the umbral dots emerged in the center of a sunspot, categorized into type-C (Li et al. 2021b). Umbral-dots emergence is an unfrequent way to form LB (as listed in Table 2), significantly less than penumbral intrusion or sunspot/umbra merging (Li et al. 2021b). The spatial resolution of telescope may limit the ability of identifying faint umbral dot chains, so the number of type-C LBs might be raised if the resolution got higher in the future. As an example, Figure 4 shows temporal evolution of a type-C LB from NOAA 11946 in maps of continuum intensity and total magnetic flux, displaying the LB formation process and subsequent active region evolution. At about 10:34 UT on 2014 Jan 8, several isolated umbral dots (UDs) appear within the sunspot. Then they gradually connect with each other in few hours, growing into an extensive bright features. Finally, the extensive structures connect to penumbrae and evolve into an LB at 02:46 UT on 2014 Jan 9. This LB also obviously corresponds to WFL as shown in the bottom panels in Figure 4 (marked by the red arrows), due to flux systems splitting. As shown in 4-6 columns of Figure 4, after the type-C LB formed, the sunspot becomes disorganized in the following days. In the bottom panel, the normalized curve for total flux evolution monotonously decrease. That is to say, the AR is in the decaying phase.

### 3.5. *Appearing Dates of LBs Vs. Peak Dates of ARs*

Table 2 lists the date of LBs appearing($T_{LB}$) and the date for the located ARs to reach its peak sunspot area($T_{AR}$). Here we calculate the time difference $\delta T$ for every LB, $\delta T = T_{LB} - T_{AR}$. The average value of $\delta T$ is 1.06, -1.60, 1.82 for type-A, B, C LBs respectively, with the standard deviation of 3.27, 2.17, 1.89. A positive value of $\delta T$ means that the LB appear in the developing phase of AR, whereas a minus $\delta T$ in the decaying phase. The most standard deviation means that type-A LBs distributes wider than others. Figure 6 exhibits $\delta T$ for the three kinds of LBs. The dashed lines show the average values. The distributions of $\delta T$ are clearly displayed in the figure, implying the temporal relation of LBs formation to ARs evolution phase. Type-A LBs trend to take shape in the decaying phase or around approaching the peak. Type-B LBs are more likely to be formed in the developing phase of sunspots. Type-C LBs are mostly formed in the decaying phase of flux systems.

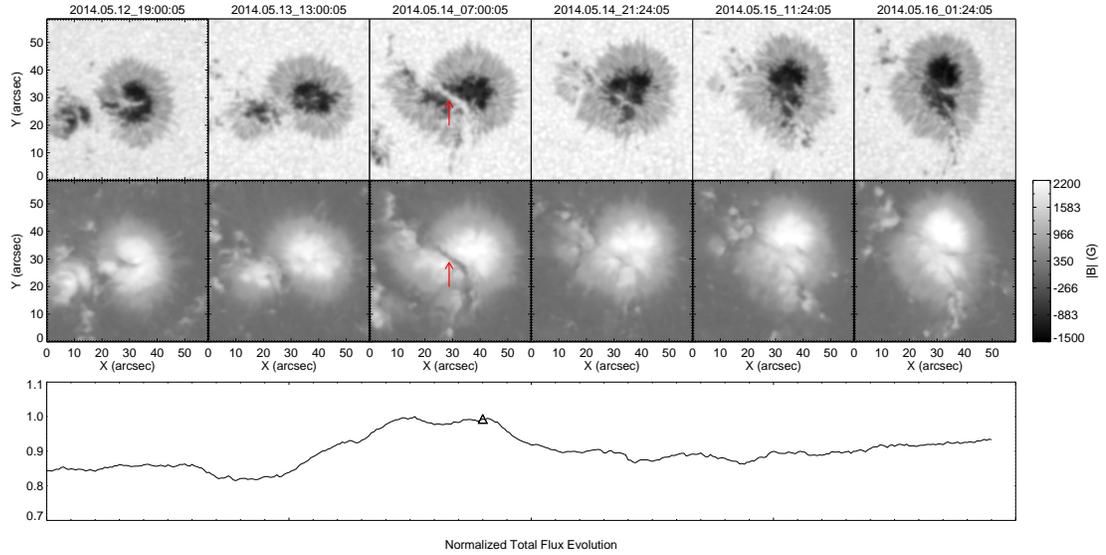

Movie 2

**Figure 2.** The same as figure 1 but for a type-B (sunspot/umbra merging) LB from NOAA 12060. (An animation of this figure is available, in the same FOV, from 2014.05.12_19:00:05 to 2014.05.17_14:12:05, showing the evolution of the LB and sited sunspot in continuum wavelength and magnetogram.)

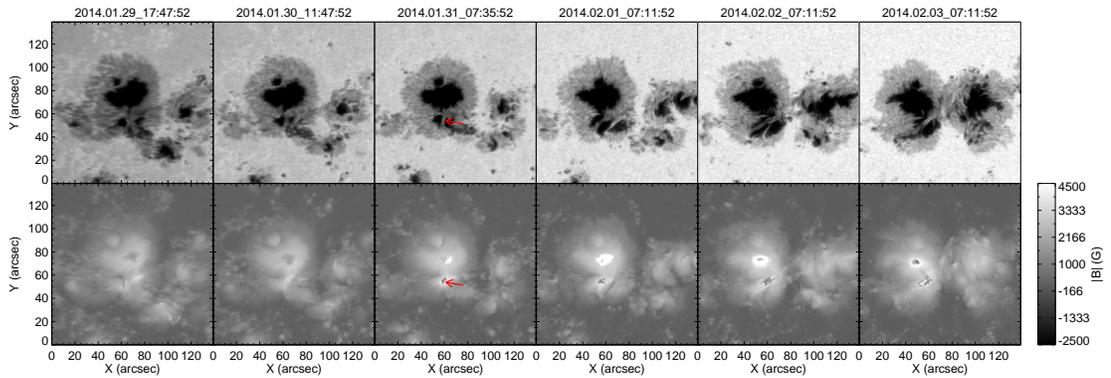

Movie 3

**Figure 3.** The formation process and subsequent AR evolution for a special type-B LB in $\delta$ sunspot from NOAA 11967. (An animation of this figure is available, in the same FOV, from 2014.01.29_10:35:52 to 2014.02.05_01:35:52, showing the evolution of the LB and sited sunspot in continuum wavelength and magnetogram.)

## 4. CONCLUSIONS AND DISCUSSION

A few studies suggests the formation mechanisms of LBs, for example, field-free hot plasma intruding, large-scale flux emerging, or inward motion of umbral dots (Rimmele 2004; Katsukawa et al. 2007; Louis et al. 2020). Based on statistics of observation samples, formation processes of LBs have been categorized into three goups: penumbral intrusion (type-A), sunspot/umbra merging (type-B) and umbral-dot emergence (type-C); (Li et al. 2021b). Li et al. (2021b) gives observational details on the formation processes of 144 identified LBs in 2014, the peak year of 24th solar cycle. Here we just employ therein 109 LBs whose length is not less than 8 arcses, to match better with the large-scale ARs by excluding small-scale LBs. We would caution that the number of type-C cases is partly limited by spatial resolution of telescope, that would be promoted with the improvement of resolution. It is worth to pay more attention to the relationship between LBs and ARs, in order to understand the formation mechanisms of LBs at the scale of flux systems. All the three types of LBs correspond to the WFLs well, except for that in special $\delta$ sunspot (for example, Figure 3). It also could obliquely explain why $\delta$ sunspots complicated and active. For type-A and type-C LB, the WFL is associated with the flux system split. For type-B LB, the WFL declining is caused by the flux system collision.

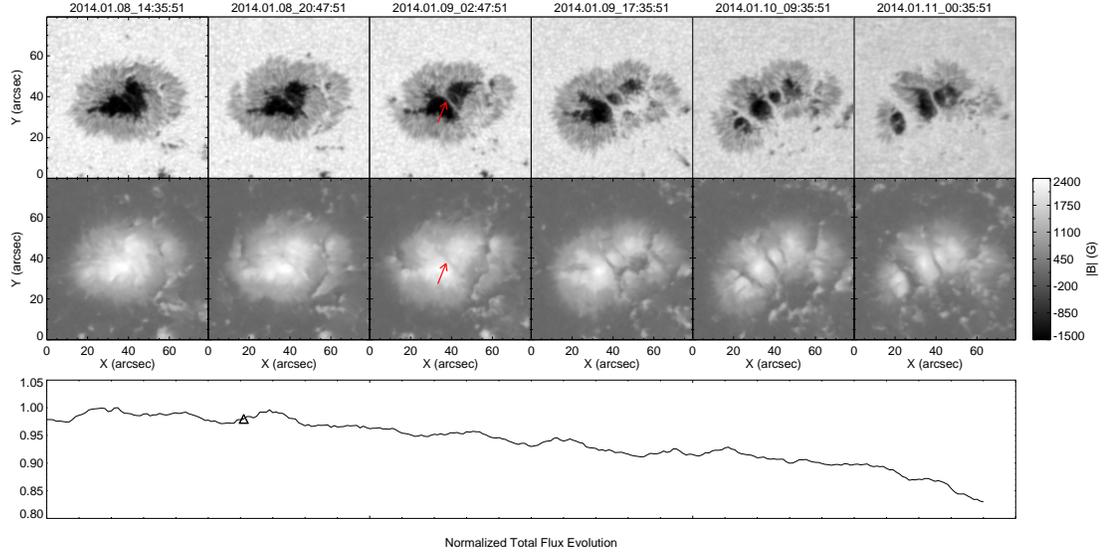

Movie 4

**Figure 4.** The same as figure 1 but for a type-C (umbral-dot emergence) LB from NOAA 11946. (An animation of this figure is available, in the same FOV, from 2014.01.08_08:35:51 to 2014.01.12_04:47:51, showing the evolution of the LB and sited sunspot in continuum wavelength and magnetogram.)

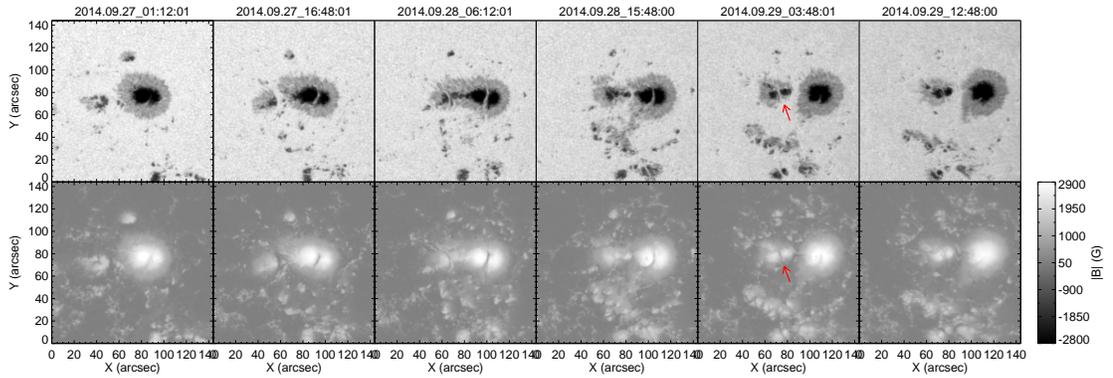

Movie 5

**Figure 5.** The formation process and subsequent AR evolution for a type-B LB from NOAA 12172. (An animation of this figure is available, in the same FOV, from 2014.09.26_03:48:01 to 2014.10.02_00:00:00, showing the evolution of the LB and sited sunspot in continuum wavelength and magnetogram.)

Sunspot/umbra merging LBs (Type-B) are formed by collisions of flux systems and more likely to be formed in the developing phase of sunspots/active regions. Penumbral intrusion LBs (Type-A) and umbral-dots emergence LBs (Type-C) are facilitated by splits of flux systems and trend to take shape in the decaying phase or epilogue of the developing phase. As shown in Figures 1 and 4, the sunspots split and decay after the type-A and type-C LBs formed. And, the total magnetic flux evolution curve is monotone decreasing, indicating decaying of ARs. In Figure 2, the umbrae merge into new one and keep steady after the type-B LB appearing. The total magnetic flux increased before the LB formed, and did not monotonely decrease substantially after that. For $\delta$ spot regions (Figure 3), the magnetic strength in LBs might be stronger than that in umbrae (Livingston et al. 2006; Wang et al. 2018a; Castellanos Durán et al. 2020). Such super strong fields of the order over 4 kG could only be caused by convergence of flux systems, but not by splitting. Special LBs within the $\delta$ sunspot appear along the polarity inversion line, with very strong and sheared horizontal magnetic fields (Livingston et al. 2006; Toriumi & Hotta 2019; Castellanos Durán et al. 2020; Lozitsky et al. 2022). A positive value of $\delta T$, the time difference between the LBs appearing and the ARs reaching peak sunspot area, means that the LB appear in the developing phase of AR, whereas a minus $\delta T$ in the decaying phase. The most standard deviation means that type-A LBs distributes wider than others. Figure 6 clearly display the relationship between LBs formation and ARs evolution phase statistically. Type-A LBs trend to form in the decaying phase or nearly approaching the peak. Type-B LBs prefer to take shape in the developing phase. Type-C LBs are mostly formed in the decaying phase of ARs.

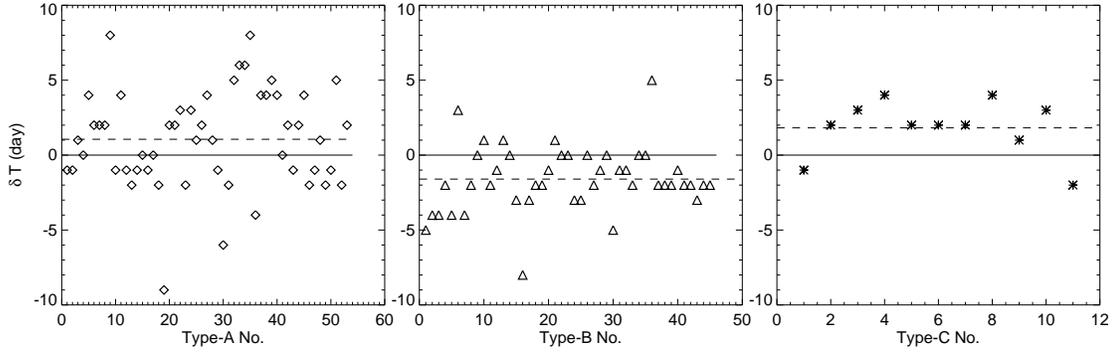

**Figure 6.** $\delta$T between LBs appearing and the located ARs reaching the peak sunspot area. These three panels are for Type A, B and C LBs respectively. The dashed lines represent the average $\delta$T for each type.

There could be multiple kinds of LBs in an AR. Listed in the 7th column of Table 2, the evolution of LBs are presented in chronological order. Generally, for multiple LB types in one AR, type-B LB is expected appearing prior to other two types, for example AR 11944, 11974, 11991, 12049, 12085, 12193, 12230, and 12241 (Table 2). For AR 11944, three type-B LBs arise several days earlier than other types of LBs in the early phase of the active region (Table 1). Of course, there could be exceptional cases, for example AR 11967, 12021, 12060, 12172. In AR 12172, when a part of umbra left from a decaying sunspot meet and merge with a newly emerging small sunspot, a type-B LB (marked by the red arrows) arises after other LBs (see in Figure 5). This AR with uncommon type evolution of LBs is complex and active, consistent the statistical conclusion that LBs are more frequently formed in ARs with complicated and multipolar magnetic fields (Li et al. 2021b).

This work was supported by the National Natural Science Foundation of China (NSFC) grants 12203054, Project Supported by the Specialized Research Fund for State Key Laboratories (CAS), and Sichuan Science and Technology Program 2023NSFSC1349. HMI is an instrument onboard the Solar Dynamics Observatory, a mission for NASA's Living With a Star program.
The authors declare that they have no conflicts of interest.